\documentclass[fleqn,11pt]{wlscirep}
\usepackage[utf8]{inputenc}
\usepackage[T1]{fontenc}
\usepackage{bm}

\newcommand{\apjs}{ApJS}
\newcommand{\apj}{ApJ}
\newcommand{\apjl}{ApJ}
\newcommand{\mnras}{MNRAS}
\newcommand{\aap}{AA}
\newcommand{\aj}{AJ}
\newcommand{\nat}{Nature}
 
\newcommand{\araa}{ARAA}

\title{Star-formation in cosmic-dawn galaxies}

\author[1,2,*]{Sandro Tacchella}
\affil[1]{Kavli Institute for Cosmology, University of Cambridge, Madingley Road, Cambridge, CB3 0HA, UK}
\affil[2]{Cavendish Laboratory, University of Cambridge, 19 JJ Thomson Avenue, Cambridge, CB3 0HE, UK}

\affil[*]{e-mail: st578@cam.ac.uk}

\begin{abstract}
In the first two years of operation JWST has delivered key new insights into the formation and evolution of galaxies in the early Universe. By combining imaging with spectroscopy, we discovered and characterised the first generation of galaxies, probing the Universe at an age of 300 million years. While the current JWST observations confirm the overall cosmological framework and the paradigm of galaxy formation, there are also surprises, including large abundances of bright galaxies and accreting black holes in the early Universe. These observations, together with detailed measurements of the stellar populations and morphological structure, will help us to develop in the coming years a more refined understanding of the baryonic physics (including star formation and feedback processes) that leads to the formation of mature systems at later epochs, including our own Milky Way galaxy.
\end{abstract}
\begin{document}

\flushbottom
\maketitle

\thispagestyle{empty}



\section*{Cosmological structure formation}

Cosmology builds on the fundamental assumption of the cosmological principle: on a large scale, the Universe is both homogeneous and isotropic. Together with general relativity as the theory for gravity, we can then relate the evolution of the Universe to its constituents and energy density. The standard cosmological model (``Hot Big Bang model'') makes accurate and testable hypotheses in several areas, including ($i$) the expansion of the Universe (Hubble–Lema\^{\i}tre law); ($ii$) the origin of the cosmic microwave background (CMB) radiation; ($iii$) the nucleosynthesis of the light elements (including deuterium, helium, and lithium); and ($iv$) the formation of galaxies and large-scale structure. The standard cosmological model has three major components, including a cosmological constant denoted by $\Lambda$ associated with dark energy, cold dark matter (CDM), and ordinary, baryonic matter. It is therefore referred to as the $\Lambda$CDM model. 

While the nature of dark matter and dark energy are still a mystery, the $\Lambda$CDM model together with detailed measurements of the CMB radiation build the foundation for the initial conditions for galaxy formation and evolution. The CMB was emitted 380,000 years after the Big Bang (corresponding to a redshift of $z\approx1,100$), containing an imprint of the density fluctuations in dark matter and the baryon-photon plasma. The baryonic fluctuations, suffering from acoustic oscillations and diffusion damping, were overall weaker than the dark matter fluctuations. These dark matter fluctuations grew with cosmic time throughout the dark ages ($z\approx1,100\rightarrow30$), forming virialized dark matter haloes. Baryons, in particular hydrogen gas, fell into these dark matter haloes and were able to cool, thereby forming molecular hydrogen and the first stars. 

The first stars -- free of any metals (Pop III) -- form around 100 million years after the Big Bang (redshift of $z\approx30$), thereby ending the so-called cosmic dark ages. The high-energy photons emitted by these stars penetrate the atomic medium around them, generating Ly$\alpha$ photons that couple the 21-cm spin and kinetic temperatures. These Ly$\alpha$ spheres are visible in the 21-cm against the CMB. Therefore, global 21-cm experiments such as REACH\cite{de-lera-acedo22} or SARAS\cite{singh18} (in addition to future gravitational wave signals) are the most promising avenues to indirectly detect the formation of the first stars. While key questions regarding the timing and physics of the first stars remain open\cite{klessen23}, massive Pop III stars of several $10^5~\mathrm{M}_{\odot}$ could form and give rise to the super-massive black holes observed in high-redshift quasars. Furthermore, it is clear that the feedback from massive Pop III stars plays a central role in regulating subsequent star formation (metal-enrich Pop II stars).

\section*{A simple model for star formation in dark matter haloes}

\begin{figure}
\centering
\includegraphics[width=\linewidth]{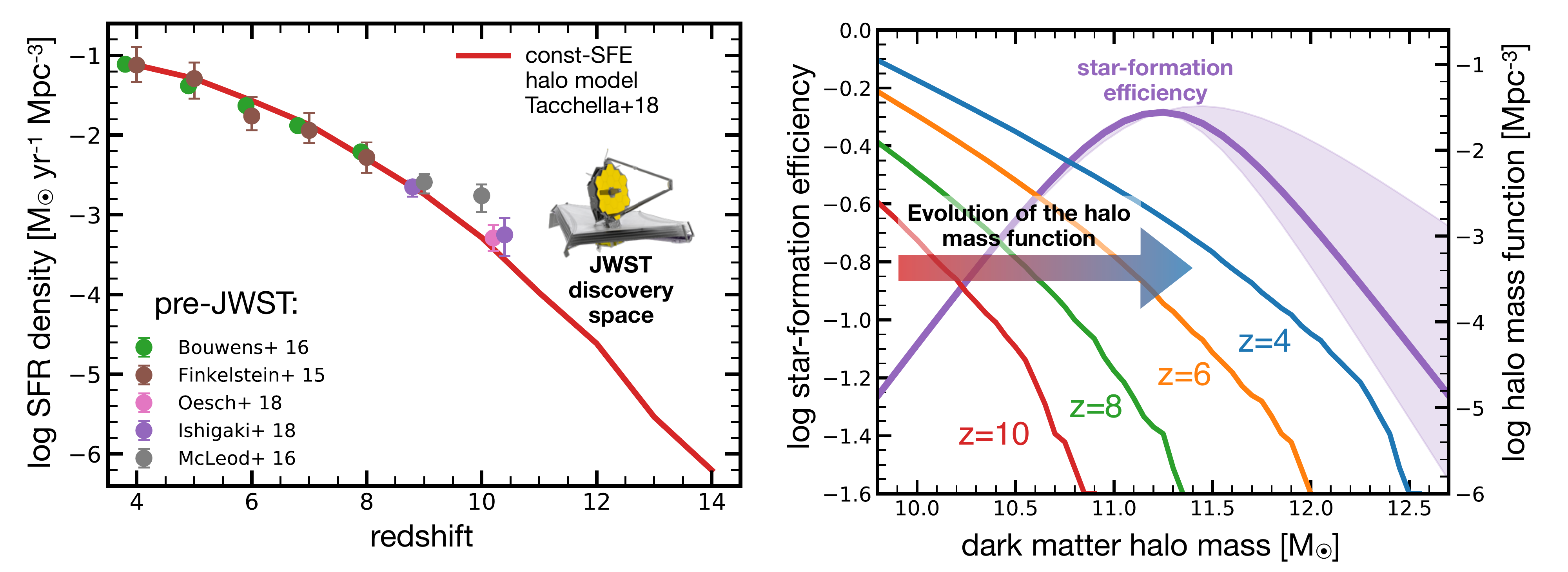}
\caption{\textit{Left:} the cosmic SFR density as a function of redshift, where redshift $z=4$ and $z=14$ correspond to 1.5 and 0.3 billion years after the Big Bang, respectively. The data points indicate the observational constraints pre-JWST, highlighting the large discovery space ($z>10$) of JWST. The solid red line shows the simple halo model, where the star-formation efficiency (SFE) is assumed to be constant\cite{tacchella13, tacchella18}. We find that that such a model predicts a rapid increase of the cosmic SFR density from $z=14$ to $z=4$ (by 5 orders of magnitude over $\sim1$ Gyr in cosmic time). \textit{Right:} the SFE and halo mass function as a function of dark matter halo mass. The reason for the strong increase of the cosmic SFR density is the rapid evolution of the halo mass function, where the number of haloes that host galaxies with a high SFE increases rapidly from $z=10$ and $z=4$.}
\label{fig:halo_model}
\end{figure}

The formation of galaxies is expected to be in full swing $180-270$ million years after the Big Bang ($z\approx15-20$). While the paradigm of galaxy formation is well established (formation of virialized dark matter haloes $\rightarrow$ cooling of gas $\rightarrow$ feedback to prevent over-cooling), star formation and the associated feedback processes from stars (such as radiation pressure, winds and supernovae explosions) and from super-massive black holes is not well understood. We can build a simple model of early galaxies\cite{tacchella13, mason15, tacchella18, harikane22_gold} by assuming that the star-formation rate (SFR) of a galaxy is directly proportional to the supply of gas. Specifically, we assume the SFR of a dark matter halo of mass $M_h$ at redshift $z$ is
\begin{equation}
    \mathrm{SFR}(M_h, z) = \varepsilon(M_h) f_b \frac{dM_h}{dt},
\end{equation}
where $\varepsilon(M_h)$ is the star-formation efficiency (SFE), $f_b$ is the cosmic baryon fraction ($f_b=0.17$), and $dM_h/dt$ is the dark matter accretion rate, which depends on $M_h$ and $z$. Here, we have assumed that the gas accretion rate is directly proportional to the dark matter accretion rate. Importantly, we further assume that the SFE $\varepsilon(M_h)$ only depends on halo mass, i.e., there is no dependency on cosmic time. Hence, this simple model can be referred to as the ``constant-SFE'' model. 

The key physics about star formation in this model is encapsulated in the SFE $\varepsilon(M_h)$. While numerical simulations and semi-analytical models try to motivate $\varepsilon(M_h)$ (or a similar function) with a description for star formation and a range of feedback processes, empirical models such as this constant-SFE model calibrate $\varepsilon(M_h)$ via observations. Using the ultra-violet (UV) luminosity function at $z=4$, this model is then able to reproduce a wide range of observations across cosmic time, including the cosmic SFR density (left panel of Fig.~\ref{fig:halo_model}), galaxy stellar mass function, star-forming main sequence, and mass-metallicity relation\cite{tacchella18}. 

The strong increase in the cosmic SFR density can be explained by studying the evolution of the halo mass function. The right panel of Fig.~\ref{fig:halo_model} shows that the halo mass function evolves rapidly from $z=10$ to $z=4$, thereby increasing the number density of galaxies that are able to host efficient star formation\cite{tacchella13}. Furthermore, the halo mass accretion rates (at fixed halo mass) also increase with cosmic time. This highlights that the primary driver of galaxy evolution across cosmic time (and in particular at $z=4-10$) is the build-up of dark matter haloes, without the need to invoke a redshift-dependent efficiency in converting gas into stars. 

Despite this strong link between dark matter haloes and galaxy formation, it is difficult to use observations of the high-redshift galaxy population to constrain different dark matter models. Specifically, within the framework of this model, changes in the dark matter model and their imprint onto the galaxy population can be mimicked by a change of the SFE\cite{khimey21}: for example, moving from a cold dark matter model to a warm dark model (which delays the formation of haloes) can be compensated by increasing the SFE in low-mass haloes. Basically, the dark matter model is degenerate with the baryonic physics. Therefore, understanding and constraining the baryonic physics of galaxy formation is crucial to also shed more light onto the nature of dark matter.

\section*{JWST: a new era of discoveries}

\begin{figure}
\centering
\includegraphics[width=\linewidth]{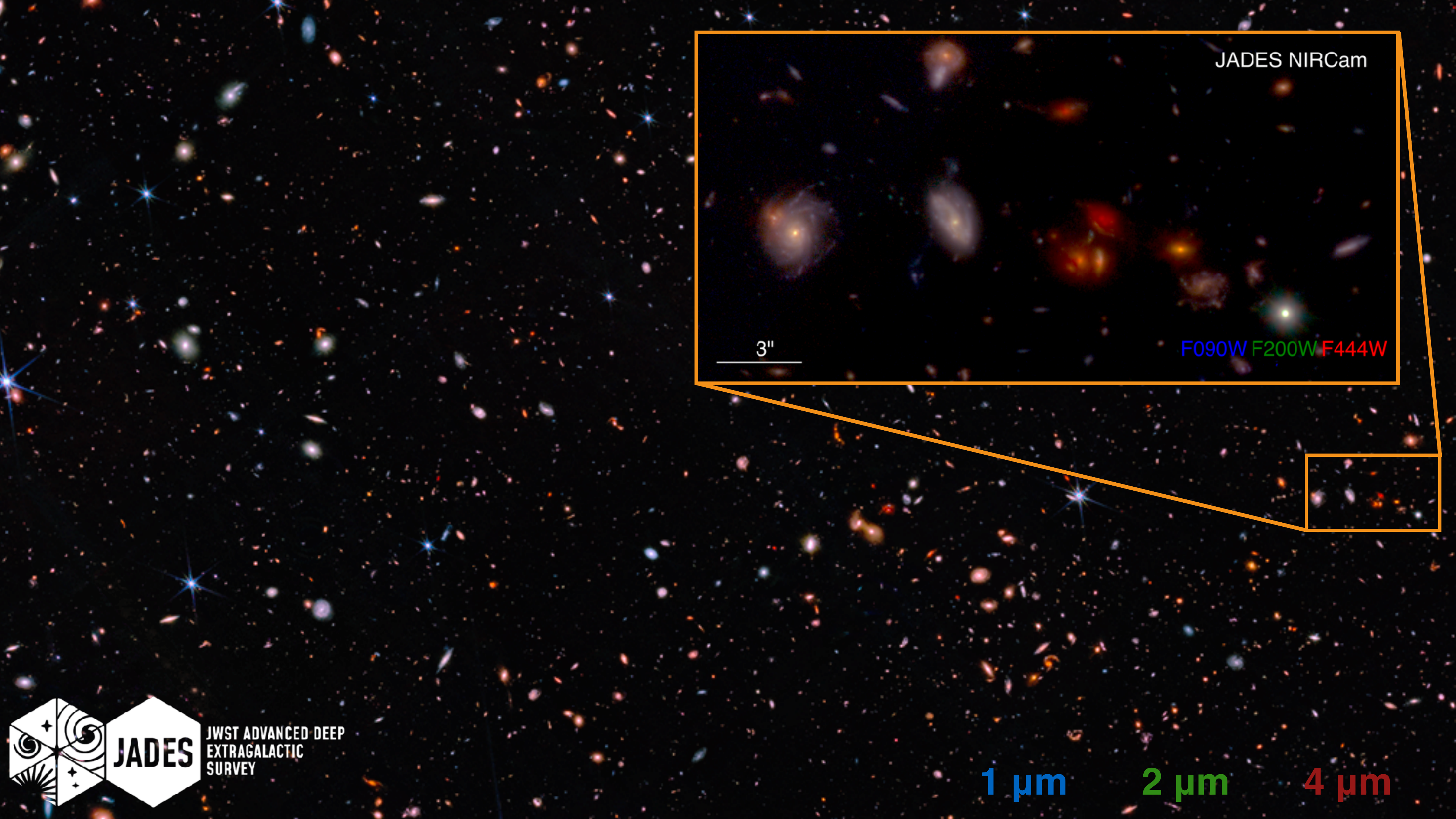}
\caption{JADES NIRCam GOODS-S imaging\cite{eisenstein23_jades}, combining F090W, F200W, and F444W filters. This image shows the great diversity of galaxies revealed in JWST images, with a wide variety of colours and morphologies.}
\label{fig:jades}
\end{figure}

Now, with the advent of JWST, we are able for the first time to directly detect the first generation of galaxies at $z\approx10-20$ and characterise their decedents in detail during the Epoch of Reionization (EoR; $z\approx6-9$). JWST, with its four science instruments (NIRCam, NIRSpec, MIRI and NIRISS), can take images and spectroscopy in the near-infrared (NIR) at wavelengths $0.6-28~\mu$m. For performing detailed measurements of the stellar populations and morphology of early galaxies, NIRCam and NIRSpec -- working at $0.6-5~\mu$m -- are the key instruments, probing the rest-frame UV and optical (out to $z\approx9$) properties of galaxies. While NIRCam provides the deepest image ever taken of the cosmos and allows us to discover the faintest structures (Fig.~\ref{fig:jades}), NIRSpec's multi-object spectrograph provides us with the first NIR high-resolution ($R>1000$) spectra from space. 

\section*{Finding and characterising the first galaxies}

One of the key science goals of JWST is the finding and characterisation of the first generation of galaxies. While the Hubble and Spitzer Space Telescopes provided us with samples of $z>9$ galaxy candidates, the photometric redshifts were uncertain and it was challenging to confirm their distances with spectroscopy. In the first two years of JWST operation, several extragalactic surveys have been developed and conducted with the aim to find new high-redshift galaxy candidates \cite{robertson22}, including 
the \textit{JWST Advanced Deep Extragalactic Survey}\cite{eisenstein23_jades} (JADES; PIs Rieke and L\"utzgendorf),
the \textit{JADES Origins Field}\cite{eisenstein23_jof} (JOF; PIs Eisenstein \& Maiolino),
the \textit{JWST Extragalactic Medium-band Survey}\cite{williams23_jems} (JEMS; PIs Williams, Tacchella \& Maseda),
the \textit{Prime Extragalactic Areas for Reionization Science}\cite{windhorst23} (PEARLS; PIs Windhorst \& Hammel),
the \textit{Grism Lens Amplified Survey from Space}\cite{treu23} (GLASS; PI Treu),
the \textit{Cosmic Evolution Early Release Science Survey}\cite{bagley23} (CEERS; PI Finkelstein),
the \textit{Next Generation Deep Extragalactic Exploratory Public Survey}\cite{leung23} (NGDEEP; PIs Finkelstein, Papovich \& Pirzkal),
COSMOS-Web\cite{casey23} (PIs Kartaltepe \& Casey), and
the \textit{Ultradeep NIRSpec and NIRCam ObserVations before the Epoch of Reionization}\cite{bezanson22} (UNCOVER; PIs Bezanson \& Labb\'{e}).

\begin{figure}
\centering
\includegraphics[width=\linewidth]{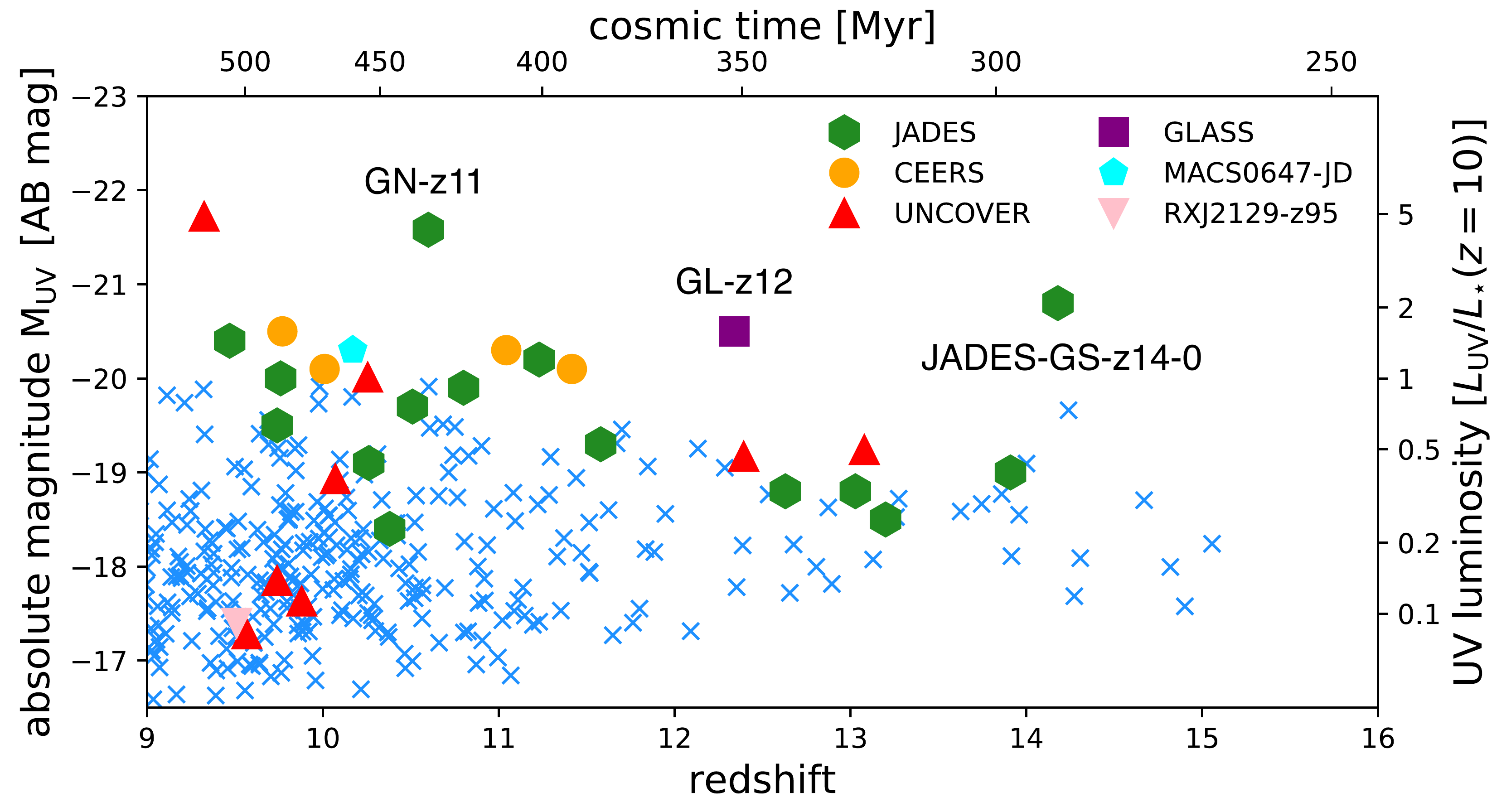}
\caption{Absolute UV magnitude versus redshift. The blue crosses mark the high-redshift galaxy candidates with photometric redshifts from the JADES survey\cite{hainline24}, while the other symbols show spectroscopically confirmed galaxies from a range of surveys\cite{curtis-lake23, arrabal-haro23, fujimoto23_uncover, hsiao23, williams23_z9}.}
\label{fig:zM}
\end{figure}

Within the first six months of JWST operation, the JADES team demonstrated the power and success of JWST (Fig.~\ref{fig:zM}). Four galaxies at $z>10$ have been spectroscopically confirmed\cite{curtis-lake23}, out of which two (with redshifts $z_{\rm spec}=13.2$ and 12.63) have been newly discovered with NIRCam imaging\cite{robertson23}, while the other two ($z_{\rm spec}=11.58$ and 10.38) have been high-redshift candidates based on Hubble Space Telescope observations\cite{ellis13, bouwens22}. These distant galaxies probe the Universe at an age of $\approx300-400$ million years after the Big Bang. This means that the light has travelled to us for nearly 13.5 billion years. Using stellar population modelling, we find the galaxies typically contain 100 million solar masses in stars, in stellar populations that are less than 100 million years old. The moderate SFRs and compact sizes of the order of 100 pc suggest elevated SFR densities ($\Sigma_{\rm SFR}\approx100~\mathrm{M}_{\odot}~\mathrm{yr}^{-1}~\mathrm{kpc}^{-2}$)\cite{robertson23}. 

With the aforementioned surveys, several groups\cite{naidu22_highz, castellano22, finkelstein24, adams24, atek23, donnan23, hainline24, harikane23_uvlf} have unveiled a significant number of galaxies beyond redshift $z\sim10$, but only a few have been spectroscopically confirmed\cite{curtis-lake23, arrabal-haro23, deugenio24_carbon, fujimoto23_zspec, wang23, castellano24}. Still, the conservative lower limits on the bright end of UV luminosity function and the integrated UV luminosity density at $z=9-12$ suggest a milder redshift evolution than expected by many theoretical models \cite{harikane23_uvlf}. Basically, there is very little evolution in the bright-end of the UV luminosity function from $z=9$ to $z=12$, challenging the aforementioned constant-SFE model and more complex numerical simulations. 

Pushing the frontier beyond $z\sim13$ is extremely challenging with the wide NIRCam filters, which have been mostly adopted in the Cycle 1 observations. This is because the broad-band spectral energy distributions (SEDs) of $z\sim15$ galaxies cannot be differentiated from lower-redshift (in particular $z\sim5$) galaxies \cite{naidu22, zavala23}. This challenge is addressed by the JOF survey\cite{eisenstein23_jof}, which uses medium bands, including NIRCam F162M, to isolate the Ly-$\alpha$ break at $z\gtrsim12$ and simultaneously control for contamination by lower-redshift line-emitting galaxies. JOF, building on-top of JADES, is the deepest imaging field observed with JWST, providing JWST images with 14 filters spanning $0.8-5\mu\mathrm{m}$, including 7 medium-band filters, and reaching total exposure times of up to 46 hours per filter. By constructing the deepest imaging ever taken at these wavelengths (reaching as deep as $\approx31.4$ AB mag), we identify a sample of eight galaxy candidates at redshifts $z=11.5-15$\cite{robertson24}. These objects show compact half-light radii of $R_{1/2}\approx50-200$ pc, stellar masses of $10^7-10^8~\mathrm{M}_{\odot}$, and SFR of $0.1-1~\mathrm{M}_{\odot}~\mathrm{yr}^{-1}$. By carefully assessing the completeness, we constrain the UV luminosity function at $z=12$, which is consistent with prior results. Furthermore, we  find that the UV luminosity density declines by a factor of $\sim2.5$ from $z=12$ to $z=14$, confirming the rather slow decline with redshift. The NIRSpec follow-up of JOF has been highly successful, with confirming two luminous galaxies at $z_{\rm spec}=14.32_{-0.20}^{+0.08}$ and $z_{\rm spec}=13.90\pm0.17$, being currently the highest-redshift systems known to us \cite{carniani24}. Recent ALMA follow-up confirmed the redshift of JADES-GS-z14-0 to be $14.1796\pm0.0007$ thanks to a detection of the [OIII]88$\mu$m, which is consistent with the candidate CIII] line detected in the NIRSpec spectrum \cite{carniani24_alma, schouws24_alma}. This detection is consistent with a gas-phase metallicity of $\sim0.1~\mathrm{Z}_{\odot}$, which is somewhat unexpected given the weakness of the UV emission lines, but can be reconciled with an escape fraction of ionising photons of $\approx\%$. 

\section*{Understanding early UV-bright galaxies}

In summary, the early studies mentioned above indicate that there is a higher abundance of bright galaxies at high redshifts than expected from theoretical models pre-JWST, including the constant-SFE model\cite{tacchella13, mason15, tacchella18}. Importantly, several of the brightest galaxies at $z>10$ have been spectroscopically confirmed (Fig.~\ref{fig:zM}), including GN-z11\cite{bunker23_gnz11, tacchella23} at $z_{\rm spec}=10.60$ with $M_{\rm UV}=-21.6$, GHZ2/GLASS-z12\cite{castellano24, zavala24} at $z_{\rm spec}=12.34$ with $M_{\rm UV}=-20.5$, and JADES-GS-z14-0 at $z_{\rm spec}=14.18$ with $M_{\rm UV}=-20.8$ \cite{carniani24, carniani24_alma, schouws24_alma}. So why is there an over-abundance of UV-bright galaxies at $z>10$? Several studies have considered physics beyond the standard $\Lambda$CDM model, including a modified primordial power spectrum \cite{padmanabhan23, parashari23}, alternative dark matter models\cite{bird23, dayal24}, or considering early dark energy \cite{shen24_ede}. However, there is also a wide range of physical interpretations related to baryonic physics to explain the tension. 

For example, we can extend the constant-SFE model by assuming that the SFE increases towards earlier cosmic times. There are models that predict a high SFE (feedback-free starbursts\cite{dekel23, li23}) at high gas densities and low metallicities -- conditions that are more prevalent at higher redshifts. Connected, it is expected that the variability (or burstiness) of star-formation increases at high redshifts\cite{tacchella20}, because the characteristic galactic dynamical timescales become too short for supernova feedback to effectively respond to gravitational collapse\cite{faucher-giguere18}. Since the halo mass function is so steep, this star-formation variability scatters galaxies in low-mass haloes up to the bright-end of the luminosity function, thereby increasing the abundance of bright galaxies\cite{mason23, sun23_bursty, shen23}. Alternatively, one could modify the conversion between the SFR and the UV luminosity by ($i$) a redshift-dependent dust evolution with basically zero dust attenuation at high redshift\cite{ferrara23, mirocha23}; ($ii$) a more top-heavy stellar initial mass function\cite{inayoshi22, yung24}; or ($iii$) contribution to the UV from non-stellar sources, including black holes\cite{tacchella23}. 

\begin{figure}
\centering
\includegraphics[width=\linewidth]{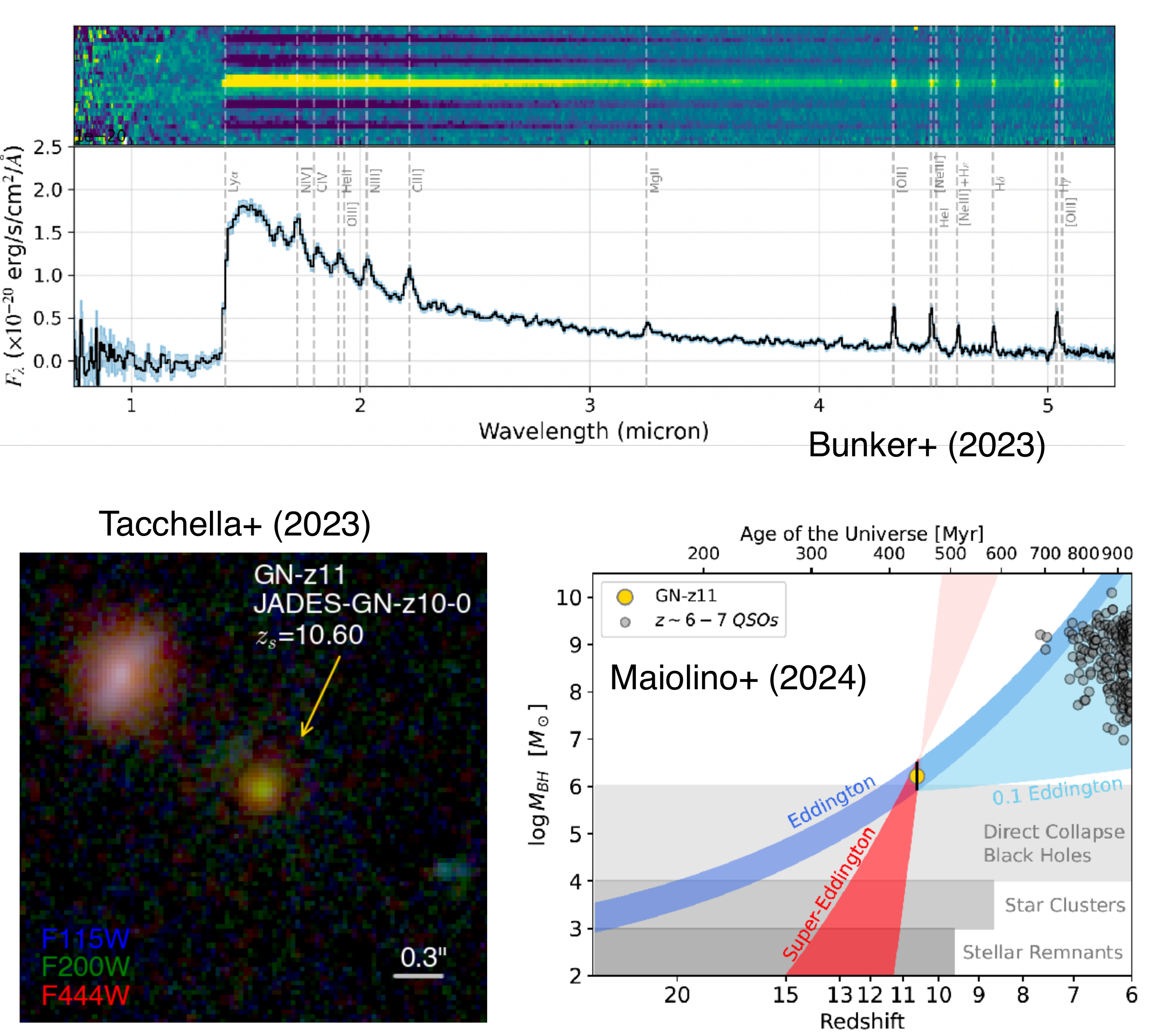}
\caption{Observations of GN-z11, currently the highest-redshift known AGN ($z_{\rm spec}=10.60$). GN-z11 is very bright for its redshift (see Fig.~\ref{fig:zM}). \textit{Top:} prism spectrum of GN-z11, showing many emission lines and a clear Ly-$\alpha$ drop\cite{bunker23_gnz11}. \textit{Bottom left:} NIRCam imaging data of GN-z11\cite{tacchella23}, showing that GN-z11 is compact and can be decomposed into a point source and extended component of size $\sim200$ pc. \textit{Bottom right:} GN-z11 hosts a black hole mass of about a million solar masses, accreting at about 5 times the Eddington rate\cite{maiolino24_gnz11}. GN-z11 lies above the local stellar-to-black hole relation, which is consistent with a scenario of either heavy seeds and/or super-Eddington accretion.}
\label{fig:gnz11}
\end{figure}

The current observations cannot rule out any of those scenarios. On the contrary, we actually find support for several of them, so it is likely that all of these processes are at work. Galaxies are indeed very compact and metal-poor at $z\sim10$, consistent with the feedback-free starburst picture. They seem to contain also very little dust. Furthermore, there are clear indications from stellar populations analyses that high-redshift galaxies have bursty star formation\cite{endsley24, looser23_pop, simmonds24, tacchella23}. While most of those systems are observed during a burst (observational bias), low-mass quiescent galaxies have also been discovered\cite{looser23, looser23_pop}, which are probably only temporally quiescent\cite{dome24} (``mini-quenching'').

Most interestingly, we can also find indication for black hole growth. A prime example of this is the galaxy GN-z11, which has been a high-redshift galaxy candidate with a grism redshift of $z_{\rm grism}=11.09$ from the Hubble Space Telescope\cite{oesch16} for many years\cite{bouwens10_nicmos}. The intriguing aspect GN-z11 is that it is extremely bright ($M_{\rm UV}=-21.6$ AB mag) for its redshift ($z>10$; see Fig.~\ref{fig:gnz11}). The JADES team has confirmed the redshift of GN-z11 to be $z_{\rm spec}=10.60$ using NIRSpec\cite{bunker23_gnz11}. The NIRCam imaging\cite{tacchella23} of GN-z11 shows that this galaxy contains a nuclear point source, which outshines the underlying rest-frame UV emission of the galaxy by factor of 3. The underlying galaxy has a size of $\sim200$ pc and contains a stellar mass of $10^{8.9}~\mathrm{M}_{\odot}$. We interpret the point source to be an accreting black hole\cite{maiolino24_gnz11}, because the NIRSpec spectrum is consistent with the one of a Broad Line Region of Active Galactic Nuclei (AGN). Assuming local virial scaling relations, we derive a black hole mass of about a million solar masses, accreting at about 5 times the Eddington rate. Similar AGNs have been found in the EoR\cite{maiolino23_bh, greene24, matthee23, harikane23, juodzbalis24}, typically lying a factor of $10-100$ times above the local stellar-to-black hole mass relation, which is consistent with a scenario of either heavy seeds and/or super-Eddington accretion.

\section*{Future characterisation of early galaxies}

JWST works! After nearly two years of operation, we have over 14 spectroscopically confirmed galaxies at $z_{\rm spec}>10$ and over $\sim400$ $z>10$ galaxy candidates. These galaxies contain roughly a 100 million solar masses in stars and are actively forming stars, doubling their stellar mass every few tens of millions of years. The compact sizes ($\sim100$ pc) imply high SFR surface densities. In the future, it will be important to connect this first generation of galaxies with their descendants in the EoR, for which we can access the powerful rest-frame optical diagnostics. This will allow us to probe the stellar population and morphological structure in detail. For example, spatially resolved studies\cite{baker23, baggen23} indicate that EoR galaxies have a compact core and are growing inside-out. Overall, it seems that the central regions of galaxies assemble very early (within the first billion years after the Big Bang), given the compact sizes of galaxies at $z>10$\cite{robertson23, robertson24} and the overly massive black holes during EoR\cite{maiolino24_gnz11, maiolino23_bh, harikane23}. 

\bibliographystyle{mnras} 

\section*{Acknowledgements}
I would like to thank the JWST instrument teams at the European Space Agency and the Space Telescope Science Institute for the support that made all these discoveries possible. I thank William Baker and Charlotte Simmonds for providing useful feedback to this manuscript.

\end{document}